\documentclass[]{aa}   
\usepackage{graphics}
\usepackage{psfrag}
\usepackage{latexsym}

\begin{document}

\title{About the morphology of dwarf spheroidal galaxies and their dark matter content}
\titlerunning{Morphology and dark matter in dSphs}
\authorrunning{Walcher et al.}

\author{C.J. Walcher\inst{1} \and J.W. Fried\inst{1} \and A. Burkert\inst{1} \and R.\ S.\ Klessen\inst{2,3}}

\institute{Max-Planck-Institut f\"ur Astronomie, K\"onigstuhl 17, 69117 Heidelberg, Germany \and  Astrophysikalisches Institut Potsdam, An der Sternwarte 16, 14482 Potsdam, Germany \and Sterrewacht Leiden, P.O.\ Box 9513, 2300 RA Leiden, The Netherlands}

\offprints{C.J. Walcher}
\mail{walcher@mpia.de}

\date{Received date/ Accepted date}

\abstract{The morphological properties of the Carina, Sculptor and Fornax dwarfs are investigated using new wide field data with a total area of 29 square degrees. The stellar density maps are derived, hinting that Sculptor possesses tidal tails indicating interaction with the Milky Way. Contrary to previous studies we cannot find any sign of breaks in the density profiles for the Carina and Fornax dwarfs. The possible existence of tidal tails in Sculptor and of King limiting radii in Fornax and Carina are used to derive global M/L ratios, without using kinematic data. By matching those M/L ratios to kinematically derived values we are able to constrain the orbital parameters of the three dwarfs. Fornax cannot have M/L smaller than 3 and must be close to its perigalacticon now. The other extreme is Sculptor that needs to be on an orbit with an eccentricity bigger than 0.5 to be able to form tidal tails despite its kinematic M/L.}

\maketitle

\section{Introduction} 
\label{s:intro}

Dwarf spheroidal galaxies (dSphs) are generally characterized by low masses (smaller than $10^{8}$ M$_{\sun}$), low surface brightnesses, low central concentrations, old populations, and the absence of interstellar gas. Due to their intrinsic faintness, their study has been and still is very difficult.

Despite these apparent ``signs of insignificance'', dSphs play an important role in Cold Dark Matter (CDM) scenarios, where larger galaxies are built out of smaller building blocks. Comparatively high velocity dispersions are found in dSphs which could be the signature of a fairly large amount of dark matter inside the visible radius. Indeed, mass to light (M/L) ratios derived from virial equilibrium can be as high as 100 for some dSphs (Mateo \cite{mateo}). Dwarf spheroidals would thus represent the low mass end of the mass function of dark matter halos. Yet the observed number of dwarf galaxies is by two orders of magnitudes too small compared to the prediction of Cold Dark Matter simulations (Klypin et al. \cite{klypin}, Moore et al. \cite{moore}). This is a major challenge for CDM scenarios, which are very successful in explaining the large scale structure of the universe. There have been several propositions to alleviate this problem, which mainly deal with feedback effects or changes to CDM theory. Recently, D'Onghia et al. (\cite{elena}) have shown that self-interacting dark matter fails to solve this abundance discrepancy. More conservative solutions include feedback and heating by an ionizing UV background, thereby separating the dark from the luminous matter. Along this line of thought Stoehr et al. (\cite{stoehr}) identified the Galactic satellites with the most massive satellite substructures in their CDM simulations and inferred that tidal tails and extra-tidal stars should not, then be present in most systems because they are embedded in larger dark matter halos.

On the other hand, dwarf galaxies without dark matter can be produced naturally in the course of mergers between bigger galaxies (Barnes \cite{barnes}); they form as clumps in tidal tails. In the absence of dark matter, the Galactic dSphs will be disturbed by the tides from the gravitational potential of the Milky Way. Their large stellar velocity dispersions would then either be the result of the line-of-sight extension of the galaxy or denote unbound stars following the same orbits for a number of orbital periods after disruption of the dwarf galaxy. Then the morphology of the dSphs should be disturbed and tidal tails should be visible. Kroupa (\cite{pavel}) and Klessen \& Kroupa (\cite{kroupa}) have carried out simulations of dwarf galaxies without dark matter in the potential of a larger parent galaxy. They are able to produce galaxies with similar features as those observed for some dSphs. A conclusive test to distinguish between the tidal and the dark matter model could come from analysis of the horizontal branch morphology of stars of the dSph galaxy in the color magnitude diagram (Klessen \& Zhao \cite{klessen}). To what extend the presence of tidal tails would rule out dark matter is not entirely clear. Mayer et al. (\cite{mayer}) state that, given a flat core for the DM halo, the production of tidal tails is possible. Yet Burkert (\cite{burkert96}) argued that ``The existence of extratidal stars in Carina, Draco and Ursa Minor would demonstrate that these dSphs cannot contain significant amounts of dark matter.''

Therefore it is of great interest to investigate the morphology of the satellites of the Milky Way. Irwin and Hatzidimitriou (\cite{IH}, hereafter IH) used star counting techniques to investigate the morphology of all eight Galactic dSphs known at that time. They found that their stellar density profiles are well fitted by an empirical King profile (King \cite{king62}) in the inner parts. Yet they find for almost all dSphs a departure from the same profile (or break in the profile) at several core radii, indicating a ``break population'' of stars. As the King profile is thought to be the density profile for a relaxed satellite in an external potential, this has often been taken as evidence for the existence of tidal tails or extended components of unknown origin.

Recently, Odenkirchen et al. (\cite{michael}) have used data from the Sloan Digital Sky Survey to investigate the morphology of the Draco dwarf. Due to the immense sky coverage and the availability of five photometric bands, they were able to carry out a more sophisticated analysis. They defined a region in colour-magnitude space holding chiefly Draco stars. By selecting only the stars in this region, they were able to reduce the ``background'' of objects - consisting of foreground stars and background galaxies - by one order of magnitude. 
They could not confirm the existence of extratidal stars claimed in IH and Piatek et al. (\cite{piatek}), they see Draco as a fully relaxed object down to their detection limit. Tidal models for Draco can conclusively be ruled out on basis of the small dispersion of its horizontal-branch stars in the color-magnitude diagram (Klessen, Grebel, \& Harbeck 2003).A similar technique was applied to the Ursa Minor dSph by Martinez-Delgado et al. (\cite{david}). They found clear indications for tidal interaction between this dwarf and the Milky Way. Majewski et al. (\cite{steve}) used even stronger cuts in colour-magnitude space to identify Red Giant Branch stars belonging directly to each galaxy. They studied seven objects, five dSphs and two globular clusters and found a break in the radial RGB number density profile for every single object.

This paper presents new wide and deep imaging of three dSphs, namely Carina, Sculptor and Fornax, aimed at searching for extratidal stars. The data is presented in section \ref{s:data}. In section \ref{s:prof} we describe the derived isopleth plots and morphological parameters for all three galaxies. In section \ref{s:discus} we discuss the results and draw our conclusions.

\section{Observations and reduction} 
\label{s:data}

We obtained V band images of the three dwarf spheroidal galaxies in the constellations of Carina, Sculptor and Fornax with the Wide Field Imager (WFI) on the MPG/ESO 2.2m telescope on La Silla. The observing run took place on eleven nights between the 17th and the 29th of September 1999. The WFI provides a mosaic of 8 CCDs with a total FOV of $32'\times 32'$. The filling factor is 96\% due to small gaps between the chips. We covered 4 square degrees for Carina, 8.5 square degrees for Fornax and 16.25 square degrees for Sculptor. To cover these large fields, we made no attempt to dither, so that the gaps between the WFI chips show up in the subsequent analysis. The typical exposure time per field is 600s, reaching a limiting magnitude of 23.5 mag on photometric nights. We also do have some I and B band images which are not complete enough to be used, but will be supplemented hopefully next September.

 The frames were bias-subtracted and flatfielded in the usual way and we applied a median filter for cosmic ray removal. Between 2 and 6 frames were taken per field, either one after another or on different nights. For each field they were aligned and coadded to reach fainter limiting magnitudes. The object catalogue was extracted from these coadded frames using the SExtractor software of E. Bertin (\cite{bertin}). The fluxes were determined in elliptical apertures around the center of each object. As the precision of the photometry is not critical for this work, we considered it unnecessary to spend the large amount of time required to do PSF-photometry.

The atmospheric extinction was measured  from Landolt (\cite{landolt}) standard star fields. Extinction coefficients where determined on a night to night basis, but as some object fields were observed two times on different nights we applied only a medium extinction correction to all frames. The mean extinction coefficient is $0.137\pm0.019$. The final error in absolute photometry for the faintest stars is about 0.3 mag, while the brighter stars (below 20th magnitude) are precise to about 0.1 mag. No correction fo possible varying galactic extinction over the fields has been made.

Using the USNO-A1.0 catalogue we determined the astrometric calibration for each coadded frame. Then the object catalogues from each field were combined to a final catalogue. We used the overlapping parts of the fields ($\sim 15$ arcmin$^2$) to correct for photometric offsets between the fields. Due to the long duration of the observing run, conditions varied from photometric to thin cirrus. Nonetheless typical offset values are lower than 0.1 mag. The way we deal with non-uniform limiting magnitudes for the isopleth maps and the density profile determination is described in the relevant paragraphs.

SExtractor also produces a classification parameter for the detected objects which was used to discriminate between galaxies and stars and thus reduce the number of background objects. This parameter runs between 0 (galaxies) and 1 (stars). While there is a very clear dichotomy above 21st magnitude, stars and galaxies are almost indistinguishable below the limiting magnitude in each frame. From the distribution of the classification parameter and eyeball inspection we adopted a cut at 0.32 uniformly for all frames. This excludes between 4\% and 10\% of the objects, depending on how many stars the dSph itself contributes to the whole count. We verified that the excluded objects are distributed uniformly over the Carina field. Unfortunately this is not true for the Sculptor and Fornax fields as the seeing varied between 1'' and 2.5'' over the run and the discrimination is naturally hampered by bad seeing. We therefore choose to include all objects in the Sculptor and Fornax isopleth maps. For a discussion of the relevant effects see next chapter.

\section{Derivation of morphological parameters and determination of the profile}
\label{s:prof}

\begin{figure}[t]
  \resizebox{\hsize}{!}{\includegraphics{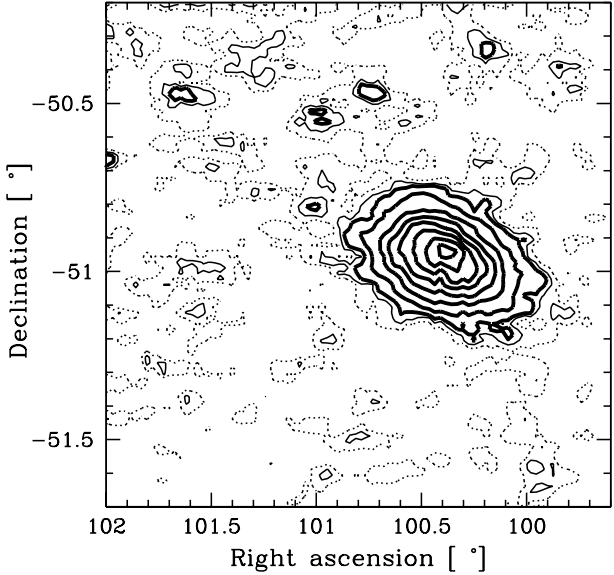}} 
  \caption[Carina contour plot]{Contour plot of the Carina dwarf spheroidal. The density levels correspond to background value (dotted line), 1$\sigma$ above that (thin solid line), 2$\sigma$, 5$\sigma$, 10$\sigma$ and so on (thick solid lines). No significant departure from the spheroidal shape can be seen. A galactic gradient can be seen from the northeastern to the southwestern corner.}
  \label{f:carcont}
\end{figure}

\begin{figure}[b]
   \resizebox{\hsize}{!}{\includegraphics{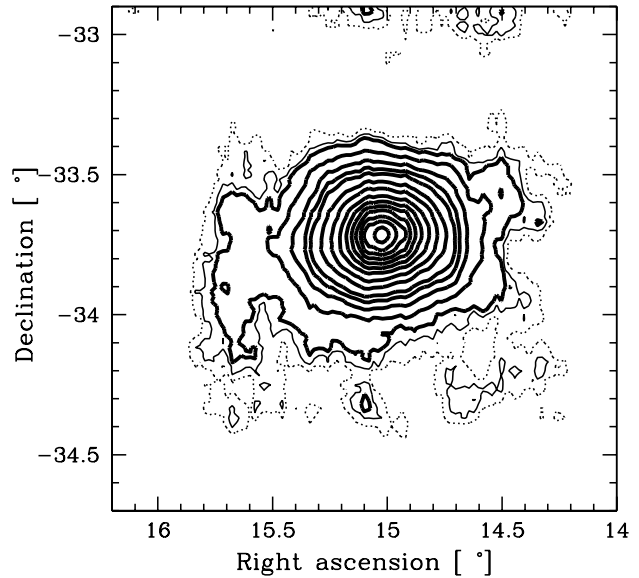}} 
   \caption[Sculptor contour plot]{Contour plot of the Sculptor dwarf spheroidal. The density levels correspond background value (dotted line), 1$\sigma$ above that (thin solid line), 2$\sigma$, 5$\sigma$, 10$\sigma$ and so on (thick solid lines). Note the increase of ellipticity with radius and the potential tidal tails.}
   \label{f:sclcont}
\end{figure}

\begin{figure}[t]   \resizebox{\hsize}{!}{\includegraphics{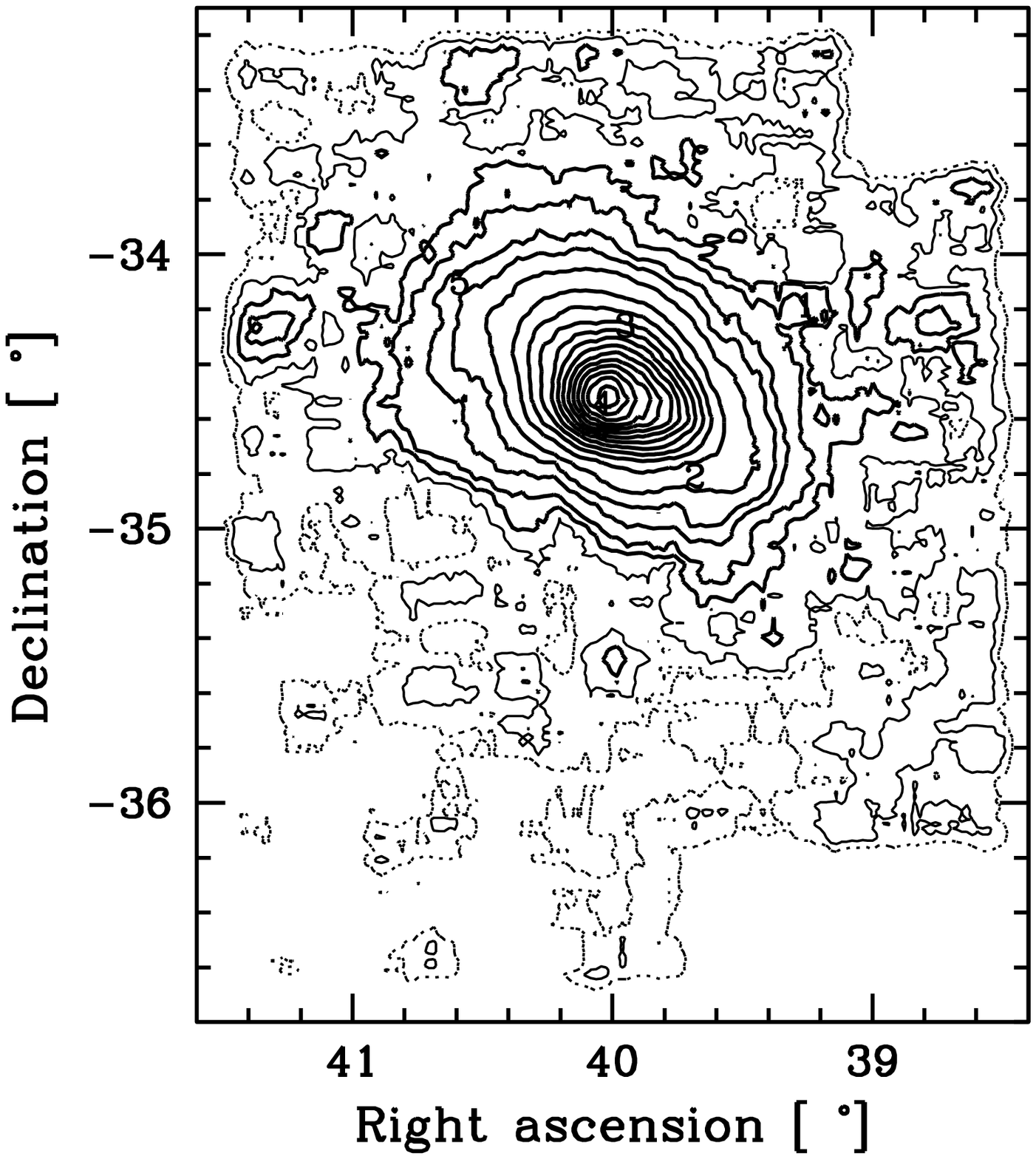}}
   \caption[Fornax contour plot]{Contour plot of the Fornax dwarf spheroidal. The density levels correspond to background value (dotted line), 1$\sigma$ above that (thin solid line), 2$\sigma$, 5$\sigma$, 10$\sigma$ and so on (thick solid lines). The inner isopleths show a steeper decline to the south-east than to the north-west. The locations of the five globular clusters are indicated by their numbers.}
   \label{f:forcont}
\end{figure}

From the final object catalogues we produced the stellar density maps shown as isopleth plots in Figures \ref{f:carcont},\ref{f:sclcont} and \ref{f:forcont}. To reduce the effects of the gaps between the WFI chips we first produced higher resolution density maps, where the gaps can be easily identified as vertical or horizontal rows with a density value of zero. We filled in these parts by taking the simple average of the surrounding pixels, excluding the ones already defined as part of the gaps. We then rebinned all density maps to have a pixel size of 1 arcmin$^2$. This procedure does not add any information to the images but prevents the eye being mislead by spurious artefacts. While we used 23.5 mag as the flux limit, some frames have a shallower limiting magnitude than others due to differing observing conditions. Their locations on the maps are pointed out in the text. As their shape is rectangular, they can hardly be confused with real structure (see for example the lower left corner of the Carina map).

The final object catalogues contain foreground Galactic stars, stars from the dwarf galaxy itself and background galaxies. These three components all have different distributions in apparent magnitude. We took advantage of this additional information by subdividing the object catalogue into magnitude bins, indexed by $i$ in the following. We then produced density maps $img_i$ in each magnitude bin. We also produced apparent magnitude histograms of a major part of the background area (called $back_i$) and one area in the galaxy center ($gal_i$). We then determined weights in the following way:

\begin{equation}
w_i^{gal} = \frac{gal_i}{gal_i + back_i}
\end{equation}

In coadding the density maps $img_i$ to the final density map $imgfin$ we used the following formula:

\begin{equation}
imgfin = (img_i - back_i) \times w_i^{gal}
\end{equation}

This improves the contrast in surface density between the center of the galaxy and the background area by $\sim 20\%$.

The density values of the contour lines in Figs. \ref{f:carcont} to \ref{f:forcont} have been chosen to be the background value (dotted line), 1$\sigma$ in the sense of statistics on a single pixel above that (thin solid line), 2$\sigma$, 5$\sigma$, 10$\sigma$ and so on (thick solid lines). The images have been smoothed to clear up the plots.

For Carina we used a particuarly low smoothing factor of 2 to be sure not to hide any signs of irregularity. We applied the star/galaxy classification. No significant departure from the spheroidal shape can be seen. In the southeast corner there is one shallower frame. A galactic gradient can be seen from the northeastern to the southwestern corner.

The Sculptor map is more difficult to interpret, because of the varying seeing conditions over the run. For consistency over the field we show the density map produced without star/galaxy discrimination. Sculptor looks very regular in the inner parts, with almost zero ellipticity. IH already noted an ellipticity increase with radius. On the outskirts two opposite annexes can be seen, where the eastern one obviously extends further to the south. These are hints but no proofs of tidal extensions, because the eastern annex actually disappears when we produce the same map taking into account the classification cut. This could either be, because the exceptional seeing on this field allows the classifier to identify more galaxies or because of a background cluster of galaxies. The reverse is true for the other annex where the seeing was exceptionally bad. This potential tail lies on a field with particularly bad seeing. Also the fields surrounding this interesting part of the density map have somewhat shallower limiting magnitude due to weather, so that we can not follow the potential tails further out. We proposed further observations to obtain a second color which will lift this uncertainty.

The same problem of seeing makes it uncertain to apply the classification also for Fornax. The differences between the maps are negligible, we therefore choose to show the map that includes the galaxies to avoid inconsistency. Fornax' innermost isopleths show a steeper decline to the east than to the west, as noticed earlier by Demers et al. (\cite{demers}). Fornax is almost as extended as the whole field as shown by Fig 7. As can also be seen from the straight line in the isophotes, the frame south of the central field is shallower, as well as two further frames in the southeastern corner. These frames were masked out in the density profile derivation below. The apparent overdensity at the lower 39.6 right ascencsion tickmark is an edge effect of the gap interpolation routine.

Centers, position angles and ellipticities were derived on the isopleth maps using the algorithm published in Bender \& M\"ollenhof (\cite{bender}) and are listed in Table \ref{t:param}. They are generally in good agreement with the values published in IH. It is worth noting that the ellipticity of Sculptor increases with radius, from 0.15 in the very center to 0.3 at a radius of about 30 arcmin. 

To determine the density profile we went back to the object catalogues. We counted all objects brighter than 23.5mag on logarithmically spaced concentric annuli of fixed position angle and ellipticity. We did not vary these parameters over radius. The background density was determined from those parts of the profile where the density reaches a constant value and then subtracted from the profile. We checked this value on a background area in the density map, both agree. We did {\bfseries not} use the weighting scheme used in the production of the density maps but did use the star/galaxy classifier for the Carina profile. We masked out the frames with a shallower limiting magnitude discussed earlier.

Though crowding is not dominant in our CCD images, we determined and applied a correction adapted to the object extraction mechanism of SExtractor. By construction the smallest separation where the detection algorithm of SExtractor resolves two stars is 1 FWHM. Therefore one star covers an area of $\pi  \cdot \mbox{FWHM}^2$ in the frame where no other star can be detected. The area covered by stars in arcmin$^2$ is therefore 
\begin{equation}
A_{covered} = \pi \cdot \mbox{FWHM}^2 \cdot n_{\star},
\end{equation}
where $n_{\star}$ is the number of stars per arcmin$^2$. Finally the number of obscured stars per square arcminute is 
\begin{equation}
\Delta n_{\star} = A_{covered} \cdot n_{\star} = \pi \cdot \mbox{FWHM}^2 \cdot n_{\star}^2 .
\end{equation}
$\Delta n_{\star}$ was added to the number density found on the raw data. For the center of Carina the correction amounts to 2.5\%; for Fornax, the most heavily crowded galaxy, this corrrection amounts to 13\% in the very center. This approach still neglects the further probability of having multiple ($>2$) stars overlap.

\begin{table}[t]
\caption[Morphological parameters]{Morphological and physical parameters as determined in this work for the Carina, Sculptor and Fornax dSphs.}
\begin{center}
\begin{tabular}{llll}
Parameter                   & Carina                & Sculptor              & Fornax                \\[0.5ex]
$\alpha$ [J2000]            & $6 \mbox{h} 41' 34''$ & $1 \mbox{h} 00' 28''$ & $2 \mbox{h} 40' 4''$  \\
$\delta$ [J2000]            & $-50^{\circ} 57'$     & $-33^{\circ} 42'$     & $-34^{\circ} 31'$     \\
PA       [$^{\circ}$]       & $64\pm 2.5$           & $98\pm 2$             & $40\pm 5$             \\  
e                           & $0.32\pm 0.04$        & $0.2\pm 0.05$         & $0.3\pm 0.05$         \\
$r_0^{emp}$ [arcmin]        & $11.96 \pm 1.5$       & $7.56 \pm 0.7$        & $13 \pm 0.15$         \\
$r_t^{emp}$ [arcmin]        & $22.54 \pm 1.4$       & $40 \pm 4$            & $89 \pm 17$           \\
$c^{emp}$                   & $0.27 \pm 0.08$       & $0.72 \pm 0.2$        & $0.83 \pm 0.1$        \\
$(\Psi/\sigma^2)^{theo}$    & 1.2                   & 2.7                   & 3.8                   \\
$r_0^{theo}$ [arcmin]       & 14                    & 10                    & 15                    \\
$r_t^{theo}$ [arcmin]       & 33.5                  & 44                    & 98                    \\
$c^{theo}$                  & 0.38                  & 0.64                  & 0.82                  \\
$L_V^{tot}$ [$10^5\,$L$_{\odot}$]   & $(3.45 \pm 0.5) $ & $(5.6 \pm 0.5)$     & $(88 \pm 0.5)$        \\
M$_{\mbox{V}}$ [mag]        & -9.0                  & -9.5                  & -12.5    \\
\end{tabular}
\label{t:param}
\end{center}
\end{table}

A King (\cite{king66}) profile was fitted to the derived density profile. The derived parameters are listed in Table \ref{t:param}. It is important to distinguish between the King \cite{king62} profile, which we also call the empirical one, and the profile as published in 1966, called theoretical profile hereafter. The empirical profile is given by the simple equation 
\begin{equation}
I(r) = k \cdot \left\{ \frac{1}{ \left( 1+(\frac{r}{r_c})^2 \right)^{\frac{1}{2}}}- \frac{1}{\left( 1+(\frac{r_t}{r_c})^2 \right)^{\frac{1}{2}}} \right\}^2.
\end{equation}
This empirical profile is usually parameterized in terms of the concentration parameter $c = \log(r_t/r_c)$ as well as the core radius $r_c$. On the other hand, following Binney and Tremaine (\cite{BT}), the theoretical profile is given by a differential equation, Poisson's equation, relating the gravitational potential $\Psi (r)$ and the dimensionless radius $r^* = r/r_0 $, where $r_0$ is some typical radius called the core radius:
\begin{eqnarray}
\lefteqn{\frac{d}{dr^*} \left((r^*)^2\frac{d \Psi}{dr^*} \right) = } \nonumber \\ & 4\pi G \rho_1(r^*)^2\left[e^{\Psi /\sigma^2}\mathit{erf}\left(\frac{\sqrt{\Psi}}{\sigma} \right) - \sqrt{\frac{4\Psi}{\pi \sigma^2}}\left(1+\frac{2\Psi}{3\sigma^2} \right) \right]
\label{eq:kingprof}
\end{eqnarray}
The dimensionless density $\rho^* = \rho/\rho_0$ is then given by the relation 
\begin{equation}
\rho^*(\Psi) = \rho_1 \left[e^{\Psi /\sigma^2}\mathit{erf}\left(\frac{\sqrt{\Psi}}{\sigma} \right) - \sqrt{\frac{4\Psi}{\pi \sigma^2}}\left(1+\frac{2\Psi}{3\sigma^2} \right) \right].
\end{equation}
The profile is then parameterized by $\psi (0)/\sigma^2$, $\rho_0$, and $r_0$. Both profiles, fitted to an observed density profile, are indistinguishable down to a density contrast of two orders of magnitude from the central density. Yet the derived tidal radii and concentration parameters differ by up to a factor of 1.5. We decided to use the theoretical King (\cite{king66}) formula, as it has a physical interpretation. This is opposite to the practice in IH and Majewski (\cite{steve}, private communications). For comparison purposes we also quote the tidal radii as determined by a fit with an empirical King (\cite{king62}) model, although the fit is somewhat worse for all three galaxies.

\begin{figure}[t]
   \resizebox{\hsize}{!}{\includegraphics{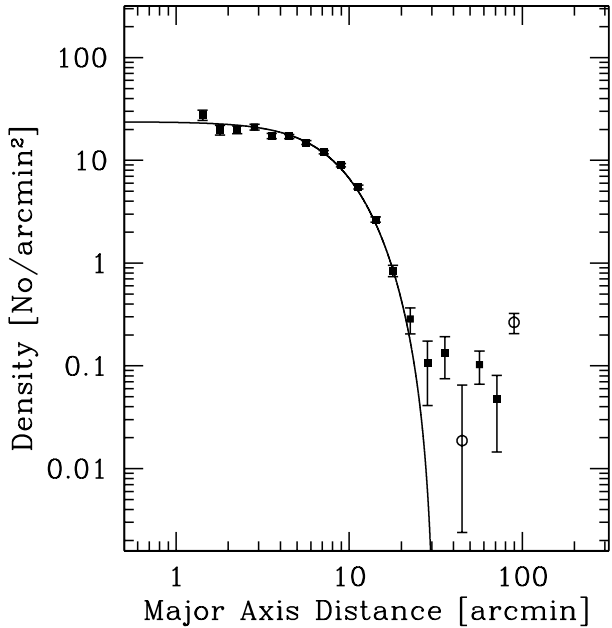}} 
   \caption[Carina profile]{Background subtracted density profile for the Carina dSph. Clearly a King profile with $r_t=31.8'$ and $c=0.35$ fits extremely well out to the limits of the data. After background subtraction negative values of the density occur which do not show up in the logarithmic plot. To illustrate these, we plotted their absolute values as open circles.}
   \label{f:carprof}
\end{figure}

For Carina the King profile fits extremely well down to the limits of the data (fig. \ref{f:carprof}). The limiting radius r$_t^{theo}$=31.8' is similar to previous studies despite the use of the somewhat different theoretical profile. The tidal radius determined from the empirical profile r$_t$ = 23.6' is 25\% smaller than the IH value (r$_t^{IH}$ = 28.8). After background subtraction negative values for the density occur where the profile drops to the background. As these negative values are not depicted in a logarithmic plot, it looks as if the profile would reach a constant positive value, yet this is not the case. The seemingly constant value of the profile is a good measure for the standard deviation of the background. To illustrate this better we also show the absolute value of the \textit{negative} values of the background as open circles. Although we show for the first time the density profile of the main sequence stars down to a density of $4 \times 10^{-3}\cdot \rho_0$ (where $\rho_0$ is the central density), we cannot confirm the break to a shallower slope found by IH and Majewski et al. (\cite{stevecar}) at a radius of 20 arcminutes. This break should occur at 1/25th of the central density. At this significance we can also rule out any unsymmetric, tidal-tail like form for this second component.

\begin{figure}[t]
   \resizebox{\hsize}{!}{\includegraphics{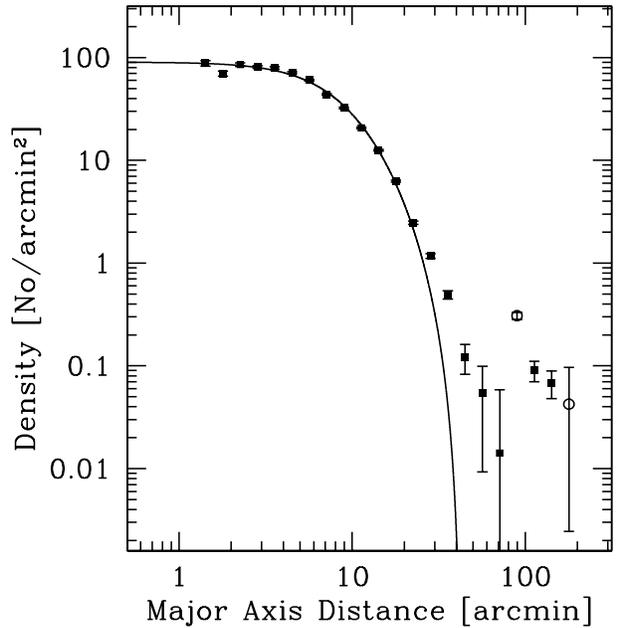}} 
   \caption[Sculptor profile]{Background subtracted density profile for the Sculptor dSph. A King profile with $r_t=70'$ and $c=0.76$ fits well out to 25'. Then the profile flattens and hits the background at about 45'. After background subtraction negative values of the density occur which do not show up in the logarithmic plot. To illustrate these, we plotted their absolute values as open circles.}
   \label{f:sclprof}
\end{figure}

For the Sculptor density profile (fig. \ref{f:sclprof}) a departure from the fitted profile can be seen clearly at 30' from the center. This break in the profile is further evidence that an extended stellar component exists. The extended component is undistinguishable from the background outside of about 45'; its interpretation is deferred to section \ref{s:discus}. The parameters determined from the King-fit again yield a smaller tidal radius and a smaller concentration parameter, compared to the IH values of 76.5' and 1.12 respectively. In this case also the values from the theoretical profile of r$_t$ = 44' and c = 0.64 are smaller than those previous determinations. We do not have an explanation for this somewhat disturbing difference. The only sizeable reduction difference lies in the crowding correction, but we would not expect this to make such a difference. We choose e=0.2 in this work as this seems to be the value in the undisturbed center of the galaxy, whereas the IH value of 0.32 might be influenced by the unresolved tidal tails. For comparison with earlier work, the corresponding values derived with an ellipticity of 0.3 are: r$_t$ = 47' and r$_c$ = 7.6'.

\begin{figure}[t]
   \resizebox{\hsize}{!}{\includegraphics{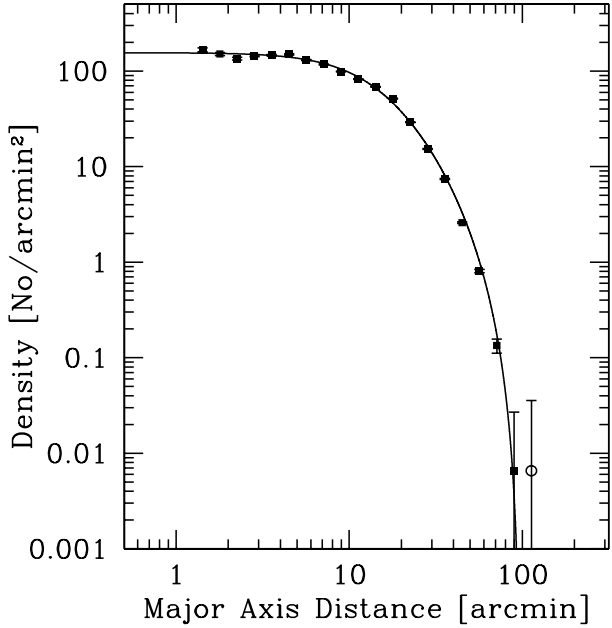}} 
   \caption[Fornax profile]{Background subtracted density profile for the Fornax dSph. Clearly a King profile with $r_t=98'$ and $c=0.82$ fits extremely well out to the limits of the data. After background subtraction negative values of the density occur which do not show up in the logarithmic plot. To illustrate these, we plotted their absolute values as open circles. The background value relies only on the two last points for this profile.}
   \label{f:forprof}
\end{figure}

The Fornax density profile (fig. \ref{f:forprof}) follows the King profile very well on to where it drops to the background noise of about 0.1 counts per arcmin$^2$. The King limiting radius r$_t$ = 98' is bigger than previously determined (IH: r$_t$=71', c=0.72) and the profile extends almost to the tidal radius derived from the King profile fit. On the other hand, our field of view is only marginally bigger than the galaxy itself. The background area is therefore not as big as in the two other cases, so a really very extended second component is not excluded by this data.

An equally acceptable fit is also obtained with Sersic profiles. They are given by 
\begin{equation} I(r) = I_e e^{-k[(\frac{r}{R_e})^{1/n} - 1]}. \end{equation}
We obtain an $R_e$ of 9', 8' and 16' while $n$ is 0.6, 0.8 and 0.7 for Carina, Sculptor and Fornax respectively. This result matches with the picture described in Caon et al. 1993 and Jerjen et al. 2000. These authors have fitted Sersic profiles to big samples of elliptical galaxies and find that the exponent in the Sersic profile follows a relation with the scale length or alternatively the absolute Magnitude. Smaller and fainter galaxies are better fitted by a larger exponent, where the largest exponents they find for small galaxies are just smaller than 2. There is no reason to rule out the Sersic profile. Nevertheless the King profile will be used as the benchmark, because it has a physical interpretation and the required cut-off. There are numerous other possible profiles, but it is beyond the scope of this paper to explore all of them.

To determine the total luminosities we first derived the luminosity function of each dwarf galaxy. To that end we normalized the distribution in apparent magnitude of all objects in a background area to the area of the whole catalogue (after applying the star/galaxy classifier only for Carina). We then subtracted this distribution function from the distribution function of the whole object catalogue. All stars belonging to the dwarf being at the same distance this yields the luminosity function of the dwarf galaxy. The integral over the luminosity function then gives the total luminosity  L$_V^{tot}$.

For Carina, using a distance modulus of m-M = 20.05 (Mighell \cite{mighell}), we find L$_V^{tot} = (2\pm 0.5) \times 10^5$ L$_{\odot}$). We excluded stars brighter than 18th magnitude to avoid contamination by bright foreground stars. Using our data alone we also have to impose a cut in apparent magnitude at 23.5 mag. To improve our estimate we extrapolated the luminosity functions of the three dwarfs using published HST data. The total luminosities we quote are therefore free of any assumptions regarding the shape of the luminosity function. For Carina we used two datasets, namely a ground-based but deeper one from Hurley-Keller et al. (\cite{denise}) and HST data from Mighell (\cite{mighell}). We first normalized the Hurley-Keller luminosity function to ours and then extrapolated further (until approximately 27th magnitude) by normalizing the Mighell luminosity function to the normalized Hurley-Keller function. This is necessary, because the number of stars brighter than 23rd magnitude is rather low in the HST data, so the intermediate step avoids errors due to small number statistics. The total luminosity of Carina is then L$_V^{tot} = (3.4\pm0.5) \times 10^5$ L$_{\odot}$. The quoted error tries to account for the unknown contribution from background galaxies in the HST data and for the very faint stars in Carina still missing in the integration

For the Sculptor dwarf the same approach as in Carina was necessary, because the relevant HST data set (Monkiewicz et al. \cite{jackie}) was originally designed to avoid crowding problems, so that there are again almost no stars brighter than 23rd magnitude. Yet the same authors find that the stellar content of Sculptor is similar to Fornax, so that we decided to interpolate the luminosity function with Fornax data from Buonanno et al. \cite{buonanno}. With a distance modulus of 19.3 (IH) and a limiting magnitude of again about 27mag, we derive L$_V^{tot} = (5.6\pm0.5) \times 10^5$ L$_{\odot}$ (from our data alone: (4.6$\pm0.5) \times 10^5$ L$_{\odot}$).

For Fornax the star catalogue found in Buonanno et al. (\cite{buonanno}) is sufficiently populated to be able to skip the intermediate step. Taking all stars brighter than 26th magnitude this yields a total luminosity of L$_V^{tot} = (8.7\pm0.5) \times 10^6$ L$_{\odot}$ (from our data alone: (6.7$\pm0.5)\times 10^5$ L$_{\odot}$), where $m-M = 20.68$.

We did not correct for stars fainter than 27th magnitude; our values might therefore be considered as lower limits. This is consistent with all three luminosities being smaller than those quoted in Grebel (\cite{eva}) by about 30\%.

\section{Summary of observations and dicussion}
\label{s:discus}

In this paper we present a new study on the morphological parameters of three Galactic dwarf spheroidal galaxies, namely Carina, Sculptor and Fornax. We show the stellar density distributions below the main sequence turnoffs in fields of 4, 16.25 and 8.5 square degrees, respectively. In the Sculptor dwarf we discover indications of tidal tails, but the quality of the current data does not allow to draw firm conclusions. Following the seminal paper by Irwin \& Hatzidimitriou (\cite{IH}) we expected to detect extratidal populations of some sort in all of our three objects. Yet we can rule out those populations for two of them. Together with similar findings by Odenkirchen et al. (\cite{michael}) for Draco this is conclusive proof that not all dSphs really have extended populations on the level claimed by IH. Nonetheless, considering Carina alone, it remains disturbing that although our density contrast should be amply sufficient we cannot confirm the extratidal component found by Majewski et al. (\cite{stevecar}) in their study of the RGB star population (see sections \ref{s:intro} \& \ref{s:prof}). This discrepancy is at the moment difficult to explain. It is possible that the extended component is so large that we do not reach the real background value, thereby overestimating it. Another possible explanation is related to the different distribution of differently aged populations in Carina found by various authors (Harbeck 2001, Monelli 2003) From our data alone we conclude that no direct sign of interaction with the tidal field of the Milky Way can be seen. This supports the idea that Carina might be dominated by dark matter. 

In the context of cold dark matter theory (CDM), the important piece of information to derive from the data is the mass of any dark matter halo in which the dSph galaxies may be imbedded. The usual approach is to determine a mass to light ratio (M/L) from kinematic and photometric data. There have been some doubts whether the kinematic approach actually determines a virialized mass or should rather be considered as an upper limit, as tidal forces might increase or even dominate the measured velocity dispersion (for example Klessen \& Kroupa \cite{kroupa}). We will therefore now determine a lower limit for the M/L from the morphology of the dwarfs only, using a similar approach as Faber \& Lin (\cite{faber}) and Pryor (\cite{pryor}). The King profile is strictly speaking only valid for a relaxed isothermal sphere with a cutoff, where no orbital motion of any kind is being taken into account. Nevertheless it can be generalized for an elliptical orbit. The perigalactic tidal radius can then be taken as a good approximation to the effective tidal radius (King \cite{king62}). Using this assumption Oh, Lin \& Aarseth (\cite{oh}) give the following relation between the King tidal radius of a satellite and its mass:
\begin{equation}
r_t = a \left( \frac{M_{dSph}}{M_G} \right)^{1/3} f(e)^{1/3}
\label{eq:tidalrad}
\end{equation}
where 
\begin{equation}
f(e) = \left\{ \frac{(1-e)^2}{[(1+e)^2/2e] \ln [(1+e)/(1-e)]+1} \right\}
\end{equation}
and where $e=(r_{apo}-r_{peri})/(r_{apo}+r_{peri})$ and $a$ are the eccentricity and the semimajor axis of the satellites orbit, respectively. M$_{dSph}$ and M$_G$ are the total mass of the dSph and the mass of the Galaxy inside a sphere with radius $a$ (see also Burkert \cite{burkert96}). For a logarithmic potential the mass of the Galaxy inside a radius $r$ is
\begin{equation}
M_G(r) = \frac{v_c^2 r}{G} \approx  1.1\times 10^{10} \left(\frac{r}{1 \mbox{kpc}}\right)\mbox{M}_{\odot}.
\label{eq:MWDMHalo}
\end{equation}
Combining equations \ref{eq:tidalrad} and \ref{eq:MWDMHalo} we have:
\begin{equation}
M_{dSph} = r_t^3\times\frac{1.1\times 10^{10}\mbox{M}_{\odot}}{a^3}\times\frac{a}{1\mbox{kpc}}\times f(e)^{-1}.
\label{eq:massdsph}
\end{equation}

In order to derive lower limits for M/L we choose the lower limits for the tidal radii to be the largest radii where the galaxies are still obviously regular in shape. For Carina and Fornax this is the point where the profile drops to the background, so we have r$_t^{car}$ = 20' and r$_t^{for}$ = 90'. For Sculptor we choose the break radius in the profile as the lower limit, which is around r$_t^{scl}$ = 25'. The lowest mass configuration is when the dSph is on a circular orbit. We thus derive the lowest possible limits for $M_{dSph}$ given in Table \ref{t:mtol}.

\begin{figure}[t]
   \resizebox{\hsize}{!}{\includegraphics{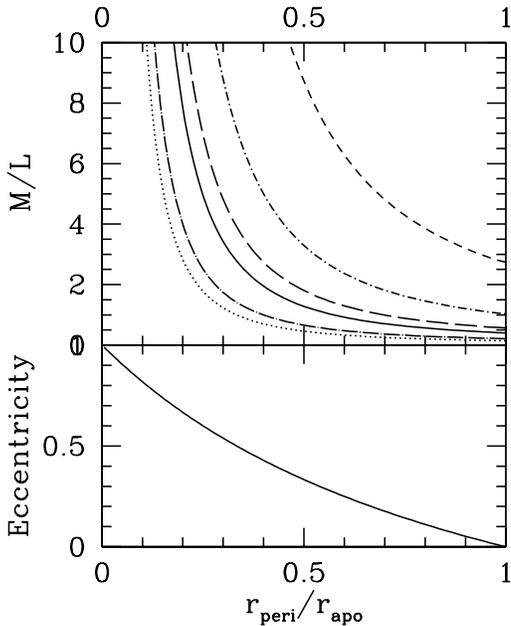}}
   \caption[Mass to light ratios]{The dependence of the lower limit for M/L on eccentricity according to equation \ref{eq:massdsph} for orbital major axes of 1 and 2 times the current distance. In the upper panel the normalized perigalactic distance is plotted against M/L. The corresponding eccentricity can be read off the lower panel. The chosen apogalactic distances are the following. Sculptor ($r_t=562$pc): $r_{apo}=88$kpc (solid line) and $r_{apo}=176$kpc (dotted). Carina($r_t=546$pc): $r_{apo}=94$kpc (long dash) and $r_{apo}=188$kpc (dot - long dash). Fornax($r_t=3600$pc): $r_{apo}=138$kpc (short dash) and $r_{apo}=276$kpc (dot - short dash).}
   \label{f:MtoL}
\end{figure}

Figure \ref{f:MtoL} illustrates how $M_{dSph}$ depends on the perigalactic distance or equivalently the eccentricity $e$ for orbital major axes $a$ of 1.5 and 2.5 times the current distance $d$ (taken from Grebel \cite{eva}). 

Alternatively Table \ref{t:mtol} also gives a ``best guess'' for $M_{dSph}$ where we choose almost randomly an eccentricity of 0.6 and an apogalacticon of 2 times the current distance. For Carina and Fornax we take the tidal radius from the King profile fit, while for Sculptor the best guess remains the radius where the first asymmetries appear.

\begin{table}[t]
\caption[Mass parameters]{Mass estimates for the Carina, Sculptor and Fornax dSphs. The quantities labelled with $^{min}$ are lower limits for the mass and M/L obtained by assuming a circular orbit. The quantities labelled $^{best}$ give a ``best guess'' based on almost random, but conservative assumptions for the orbits. Also given are the heliocentric distance, the kinematic M/L and the heliocentric velocity.}
\begin{center}
\begin{tabular}{llll}
Parameter                   & Carina           & Sculptor         & Fornax     \\[0.5ex]
Distance [kpc]              & 94               & 88               & 138         \\
M/L$_{dyn}$                 & 30               & 11               & 4.8          \\
v$_{\odot}$ [km/s]          & 225              & 100              & 50            \\
r$_t^{min}$ [pc]            & 546              & 639              & 3600           \\
M$^{min}$ [M$_{\odot}$]     & 2.0$\times 10^5$ & 3.7$\times 10^5$ & 2.7$\times 10^7$\\
M/L$^{min}$                 & 0.6              & 0.7              & 3                \\
r$_t^{best}$ [pc]           & 914              & 639              & 3928              \\
$a^{best}$ [kpc]            & 188              & 176              & 276                \\
$e^{best}$                  & 0.6              & 0.6              & 0.6                 \\
M$^{best}$ [M$_{\odot}$]    & 6$\times 10^6$   & 2$\times 10^6$   & 2$\times 10^8$       \\
M/L$^{best}$                & 17               & 3.6              & 23                    \\
\end{tabular}
\label{t:mtol}
\end{center}
\end{table}

According to Saviane et al. (2000) the main stellar population of Fornax is 5.4$\times 10^9$ years old which corresponds to an M/L in the V-band of about 4.8 according to Bruzual \& Charlot (2000). We therefore do not need any dark matter halo for Fornax. By requiring M/L to be equal to the kinematically derived M/L we can argue that Fornax must almost be in its perigalacticon. Further approach to the Galactic center would make dark matter necessary to prevent the formation of tidal tails above our detection limit.

The case is not nearly as clear-cut for Carina, because the lowest limit to M/L is so low. Yet by setting the eccentricity to 0.46 and assuming Carina to be in its apogalacticon we can easily bring M/L to be 30 as found in kinematic studies. The perigalactic distance would then be around $r_{peri} = 35$kpc. For comparison: the nearest dwarf, Sagittarius, is currently at a distance of 16 kpc from the Galactic center. Generally speaking such an eccentric orbit is in good agreement with the high Galactocentric velocity of Carina, although this radial velocity is at the same time obvious proof that Carina is not exactly in its apogalacticon now. Still, as stated by Burkert (\cite{burkert96}), a detection of tidal tails at a lower level but well inside the tidal radius adopted here, would make it difficult to accomodate a dark matter halo for Carina without invoking extreme orbital parameters.

In the case of Sculptor, to bring M/L up to the kinematic value, the eccentricity would have to be bigger than 0.5, which leads to $r_{peri} = 28$kpc. Assuming both the kinematic and the King approach to be valid we thus have to assume a highly eccentric orbit for Sculptor.

The direction of the tails compared to the direction of the Galactic center predicts a crude direction for the tangential orbital motion. The tail which is on the inner side of the main body will precede it in direction of the velocity vector, while the outer tail stays behind the main body of the dwarf. The direction of motion of Sculptor should therefore be in north-south direction. As Sculptor is very near the south Galactic pole the sign of the proper motion vector can not be inferred because it is not clear which tail is the inner one. The only proper motion measurement for Sculptor by Schweitzer et al. (\cite{schweitzer}) gives a tangential velocity vector pointing to the north-east.

One of the propositions to solve the overabundance of dark satellites is to inhibit star formation in low mass dark matter satellites by photoionisation heating and feedback. The visible dwarfs are then identified with the most massive ($\geq 10^9 \mbox{M}_{\odot}$) satellite substructures found in CDM simulations. The consequence is (citing Stoehr et al. \cite{stoehr}): ``Although there is no problem accomodating a single disrupting object like Sagittarius, it would become uncomfortable if tidal stripping were detected unambiguously in other systems.'' (see also Hayashi et al. \cite{hayashi}). Counting Sagittarius, Ursa Minor (Martinez-Delgado et al. \cite{david}) and Sculptor, there are now three satellites of the Milky Way  out of a total of 13 that are most likely currently strongly influenced by tidal forces. Note also that even our most massive galaxy, Fornax, has only $10^8$ M$_{\odot}$. We therefore conclude that our measurements do neither support nor contradict dark matter in dSphs, but the overabundance problem for Galactic satellites clearly remains unsolved.

\begin{acknowledgements}
We are grateful to Denise Hurley-Keller, Jackie Monkiewicz and Ken Mighell for making their data available to us. The anonymous referee made fruitful suggestions that led to significant improvements in the paper. CJW thanks Michael Odenkirchen for a lot of useful discussions and for lending his fitting code for the empirical King profile. RSK acknowledges support by the Emmy Noether Program of the Deutsche Forschungsgemeinschaft (DFG: KL1358/1) and travel subsidies from the Leids Kerkhoven-Bosscha Fonds (LKBF)
\end{acknowledgements}

\clearpage


\begin{thebibliography}{1999}

\bibitem[1992]{barnes} Barnes, J., Hernquist, L., 1992, Nature 360, 715B
\bibitem[1987]{bender} Bender, R., M\"ollenhof, C., 1987, A\&A 177, 71
\bibitem[1996]{bertin} Bertin, E., Arnouts, S., 1996, A\&AS 117, 393
\bibitem[1987]{BT} Binney, J., Tremaine, S., Galactic Dynamics, Princeton University Press, New Jersey (1987)
\bibitem[2000]{Bruzual} Bruzual, G., Charlot, S. 2000, in Prep.
\bibitem[1999]{buonanno} Buonanno et al., 1999, AJ 118, 1671
\bibitem[1996]{burkert96} Burkert, A., 1996, ApJ 474, L99
\bibitem[1993]{caon93} N. Caon, M. Capaccioli, M. D'Onofrio, MNRAS, 265, 1013 (1993)
\bibitem[1994]{demers} Demers, S., Irwin, M., Kunkel, W., 1994, AJ 108, 1648
\bibitem[2002]{elena} D'Onghia, E., Burkert, A., 2002, astro-ph/0206125
\bibitem[1983]{faber} Faber, S.M., Lin, D.N.C., 1983, AJ 266, L17
\bibitem[2000]{eva} Grebel, E.K. 2000, in Star Formation from the Small to the Large Scale, 33rd ESLAB Symposium, ESA SP-445, ed. F. Favata, A.A. Kaas, \& A. Wilson (Noordwijk: ESA), 87
\bibitem[2000]{daniel} Harbeck, D., et al., 2001, AJ 122, 3092
\bibitem[1998]{denise} Hurley-Keller, D., Mateo, M., Nemec, J., 1998, AJ 115, 1840
\bibitem[2002]{hayashi} Hayashi, E., Navarro, J., Taylor, J.E., Stadel, J., Quinn, T., 2002, astro-ph/0203004 
\bibitem[1995]{IH} Irwin, M., Hatzidimitriou, D., 1995, MNRAS 277, 1354
\bibitem[2000]{jerjen} H. Jerjen, B. Binggeli, K.C. Freeman, AJ, 119, 593 (2000)
\bibitem[1962]{king62} King, I., 1962, AJ 67, 471
\bibitem[1966]{king66} King, I., 1966, AJ 71, 64
\bibitem[1998]{kroupa} Klessen, R.\ S., Grebel, E., Harbeck, D., 2003, ApJ in press -- astro-ph/0302287
\bibitem[1998]{KundK} Klessen, R.\ S., Kroupa, P., 1998, ApJ 498, 143 
\bibitem[2002]{klessen} Klessen, R.\ S., Zhao, H., 2002, ApJ 566, 838
\bibitem[1999]{klypin} Klypin, A., Kravtsov, A., Valenzuela, O., Prada, F., 1999, ApJ 522, 82
\bibitem[1997]{pavel} Kroupa, P., 1997, New Aston. 2, 139
\bibitem[1992]{landolt} Landolt,A., 1992, AJ 104, 340
\bibitem[2000]{stevecar} Majewski, S., Ostheimer, J. et al., 2000, AJ 119, 760
\bibitem[2001]{steve} Majewski, S., et al., 2001, Yale Cosmology Workshop: The Shapes of Galaxies and Their Halos, astro-ph/0109492
\bibitem[2001]{david} Martinez-Delgado, D., Alonso-García, J., Aparicio, A., Gómez-Flechoso, M., 2001, ApJ 549L, 63
\bibitem[1991]{mateo91} Mateo, M., Olszewski, E., Welch, D., Fischer, P., Kunkel, W., 1991, AJ 102, 914
\bibitem[1998]{mateo} Mateo, M., 1998, ARA\&A 36, 435
\bibitem[2001]{mayer} Mayer et al., 2001, ApJ 559, 754
\bibitem[1997]{mighell} Mighell, K., 1997, AJ 114, 1458
\bibitem[1996]{moore96} Moore, B., 1996, ApJ 461, L13
\bibitem[1999]{moore} Moore, B. et al., 1999, ApJ 524, L19
\bibitem[1999]{monelli} Monelli, M. et al., 2003, AJ in press, astro-ph 0303493 
\bibitem[1999]{jackie} Monkiewicz et al., 1999, PASP 111, 1392
\bibitem[2001]{michael} Odenkirchen, M., Grebel, E., et al.,  2001, AJ 122, 2538
\bibitem[1992]{oh} Oh, K., Lin, D., Aarseth, S., 1992, ApJ 386, 506O
\bibitem[2001]{piatek} Piatek, S., Pryor, C., Armandroff, T., Olszewski, E., 2001, AJ 121, 841
\bibitem[1996]{pryor} Pryor, C., 1996, ASP Conf. Series Vol. 92, p. 424.
\bibitem[2000]{Saviane} Saviane, I., Held, E. V., Bertelli, G., 2000, A\&A 355, 56
\bibitem[1995]{schweitzer} Schweitzer, A. E., Cudworth, K. M., Majewski, S. R., Suntzeff, N. B., 1995, AJ 110, 2747
\bibitem[2002]{stoehr} Stoehr, F., White, S., Tormen, G., Springel, V., 2002, astro-ph/203342

\end{thebibliography}
\end{document}